\documentstyle[12pt,epsfig]{article}

\begin{document}
\bibliographystyle{unsrt}
\begin{titlepage}

\rightline{FNT/T-96/28}

\begin{center}
{\large\bf  On the  QED Radiator at order $\alpha^3$ }                              
\end{center}

\noindent
\begin{center}
Guido MONTAGNA$^{a,b}$, Oreste NICROSINI$^{b,a}$  
and Fulvio PICCININI$^{b,a}$ 
\end{center}

\noindent
\begin{center}
$^a$ Dipartimento di Fisica Nucleare e Teorica, 
Universit\`a di Pavia, Italy \\
$^b$ INFN, Sezione di Pavia,  Italy  
\end{center}

\begin{abstract}{\small The ${\cal O} (\alpha^3)$ leading logarithmic 
contribution to the QED radiator in the additive form is 
considered. The effect of the correction on two-fermion physics at 
LEP1 and  LEP2 is  evaluated and critically compared with the one of 
next-to-leading  ${\cal O} (\alpha^2)$ corrections. A critical comparison with existing
estimates for the LEP1 energy range is  also performed. 
The  ${\cal O} (\alpha^3)$ leading logarithmic corrections 
turn out to be relevant in view of the experimental precision already 
reached at LEP1 and foreseen at LEP2. }
\end{abstract}

\vskip 48pt \noindent
E-mail: \\
montagna@pv.infn.it \\
nicrosini@pv.infn.it, nicrosini@vxcern.cern.ch \\
piccinini@pv.infn.it \\

\vfil
\leftline{FNT/T-96/28}
\leftline{May 13, 1997}
%\leftline{\today}
\end{titlepage}

\eject
\normalsize
%%%%%%%%%%%%%%%%%%%%%%%%%%%%%%%%%%%%%%%%%%%%%%%%%%%%%%%%%%%%%%%%%%%%%%%%%

QED corrections are, as well known, an essential ingredient of precision
physics at LEP/SLC. In particular, initial-state photonic 
radiation (ISR) plays a central  r$\hat{ \hbox{\rm o}}$le in the 
determination of the energy effectively available in the
center of mass of the $e^+ e^-$ reaction. Nowadays, this effect is popularly
described by the so called Structure Functions (SF) method, 
pioneered in~\cite{sf} and subsequently developed 
in~\cite{nt,bbn,jsw,js,d3,apl,prd}. For processes of the kind $e^+ e^- \to 
\gamma Z^0 \to f \bar f$ around the $Z^0$ resonance, 
the effect of ISR is crucial in the 
precise determination of the peak cross section. Going above the $Z^0$ peak, 
ISR plays a new r$\hat{ \hbox{\rm o}}$le, namely it is responsible for the 
so called ``$Z^0$ radiative return'', i.e. the emission of hard photons such
that the energy available in the kernel reaction is brought back to the $Z^0$ 
mass.  Being the $Z^0$ radiative return caused by ISR, 
the most precise knowledge of the hard tail of the IS 
photonic spectrum is of utmost importance. 
It must be noticed that the $Z^0$ radiative return enhances the 
contribution of precisely the hard photons that reduce the centre of mass 
energy to the $Z^0$ mass. This means that most of the events are characterized by the 
fact that their invariant mass after ISR is very close to the $Z^0$ mass, or, 
in other words, they are ``LEP1-like'' events.  
Moreover, once the resonance is 
produced there is no more particular enhancement of any portion of the 
spectrum: hence final state radiation (and initial-final state interference) 
gives contributions substantially equivalent  to the ones it already 
gives at LEP1. Last, but not least, in the case of $s$-channel processes ISR 
represents a gauge invariant subset of the full set of QED corrections, and 
therefore represents a meaningful subject of investigation. 
 
In the case in which only a cut on the invariant mass of the event after ISR 
is considered, the SF method provides a very simple recipe for computing the
corrected cross section as a one-dimensional integration of the proper kernel
cross section times the so called ``radiator'' (or ``flux function''), namely
as 
\begin{equation}
\sigma (s) = \int_0^{1 - x_{cut}} dx  H (x,s) \sigma_0 \left( (1-x) s \right) , 
\end{equation}
where the radiator is defined as 
\begin{equation}
H(x,s) = \int_{1-x}^1 {{dz} \over {z}} D (z,s) D \left( {{1-x}  \over {z}} , 
s \right) , 
\label{eq:h} 
\end{equation}
$D (x,s)$ being the electron SF. 
Equation (1), with the definition (2), is a very useful tool when considering 
semianalytical calculations~\cite{tz} devoted to data analysis; actually, the
availability of simple and accurate analytical formulae is mandatory 
for the development of fast fitting programs.  

The radiators available in the literature can be classified as belonging to 
two groups, according to the kind of solution for the electron 
SF adopted: additive~\cite{sf,nt,bbn} and factorized~\cite{jsw,js,apl} 
radiators, respectively.  As a matter of fact, the additive radiators are 
extensively implemented in standard computational tools used for the two-fermion 
data analysis at LEP, for instance~\cite{tz}. From now on, we will mainly consider  
this kind of radiators. 

A typical additive radiator  consists of an exponentiated part,
taking into account soft multi-photon emission, plus finite-order
leading-lo\-ga\-ri\-thmic (LL) corrections
accounting for hard collinear bremsstrahlung 
up to ${\cal O} (\alpha^2)$, namely~\cite{sf,nt}
\begin{eqnarray}
&& H_{NT} (x,s) = \Delta_{2} \beta x^{\beta - 1} + h_1 (x,s) + h_2 (x,s) , 
\nonumber \\ 
&& h_1 (x,s) = - {1 \over 2 } \beta (2-x) , \nonumber \\
&& h_2 (x,s) = {1 \over 8 } \beta^2 \left[ (2-x) \left( 3 \ln (1-x) 
- 4 \ln  x \right) 
   - 4 {{\ln (1-x)} \over {x}} + x - 6 \right] , \nonumber \\
&& \beta = 2 \left({\alpha \over \pi } \right) \left[ L  - 1 \right] , 
\nonumber \\
&& L = \ln (s / m^2) , \nonumber \\
&& \Delta_{2} = 1 + \left( {\alpha \over \pi } \right) \delta_{1} 
   + \left( {\alpha \over \pi } \right)^2 \delta_{2} . 
\label{eq:hnt} 
\end{eqnarray}
The corrections $\delta_{1,2}$, which within the SF formalism are determined
at the LL level, can be adjusted to take into account 
process dependent soft plus virtual next-to-leading (NL)
contributions.\footnote{\footnotesize Actually, also additional pair radiation
gives contributions at the ${\cal O} (\alpha^2)$; it can be accounted for by
properly redefining the constant $\delta_{2}$ for the soft plus virtual part,
and adding a proper contribution to the hard part of the 
radiator~\cite{kkks}.} 
For  processes of the kind $e^+ e^- \to \gamma, Z^0 \to f \bar f$, 
they are known~\cite{nt} in such a way that the radiator of eq.~(3) reproduces  
the exact ${\cal O} (\alpha^2)$ soft plus virtual 
perturbative results~\cite{burgers}.  

For such processes, 
a more accurate form of the additive radiator has also been derived~\cite{berends},
taking into account ${\cal O} (\alpha^2)$ NL  hard-photon
corrections, in such a way that the full ${\cal O} (\alpha^2)$ perturbative
calculation~\cite{a2b} is reproduced, namely
\begin{equation}
H_B (x,s) = H_{NT} (x,s) - h_2 (x,s) + \delta_2^H (1-x, s) , 
\label{eq:hb}
\end{equation}
where $\delta_2^H$ is defined in  eqs.~(3.19) and (3.20) of 
ref.~\cite{berends}. The radiator (\ref{eq:hb}) differs from (\ref{eq:hnt})
by terms which turn out to be numerically negligible at the $Z^0$ peak, but can
in principle be important far from the resonance because of the well known
relevance of hard photon radiation.  Actually, in~\cite{berends} also some ansatz 
for factorized radiators can be found, which reproduce (\ref{eq:hb}) up to the 
${\cal O} (\alpha^2)$ NL corrections and take into account additional higher order 
contributions (more on this later).  

It has to be noticed that the corrections present in eq.~(\ref{eq:hb}) and
neglected in (\ref{eq:hnt}) are dominated by terms of ${\cal O} (\alpha^2 L)$, 
$L$ being the
collinear logarithm.  Being second-order NL corrections, they are in 
principle of  the same order of magnitude as the third-order LL ones.  
These last corrections are already known in the literature at the SF 
level~\cite{js,d3} and for the factorized radiator~\cite{jsw}.\footnote{\footnotesize 
The ${\cal O} (\beta^4)$ corrections to the SF are known to be at $1 \times 10^{-4}$ 
level, according to~\cite{js,d3,prz}. } 
By using the additive solution of the SF of~\cite{d3},
it is possible to derive the ${\cal O} (\beta^3)$ 
corrections to the QED additive radiator.  By combining
these  results with the ones already available, we propose the following
form for the QED additive radiator, including both NL ${\cal O} (\alpha^2 )$
and ${\cal O} (\beta^3)$ contributions:  
\begin{eqnarray}
&& H (x,s) = \Delta_{3 } \beta  x^{\beta - 1} + h_1 (x,s) 
+ \delta_2^H (1-x, s) + h_3 (x,s) , \nonumber \\
&& h_3 (x,s) = {{1} \over {3!}} \left( {\beta \over 2}\right)^3 
\left[ - {27 \over 2} + {15 \over 4 } x 
   + 4 ( 1 - {1 \over 2} x ) \left( \pi^2 - 6 \ln^2 x + 3 \hbox{\rm  Li}_2 (x) 
   \right) \right. \nonumber \\
&& \qquad \qquad + 3 \ln (1-x) \left( 7 - {6 \over x} - {3 \over  2} x \right) 
         + \ln^2 (1-x) \left( -7 + {4 \over x} + {7 \over  2} x \right) 
         \nonumber  \\
&& \qquad \qquad \left. - 6 \ln x ( 6 - x ) + 6 \ln x \ln (1-x) \left( 6 
- {4 \over x} 
   - 3 x \right) \right] , \nonumber \\
&& \Delta_{3} = \Delta_{2} + \left( {\alpha \over \pi} \right)^3 
\delta_{3} , \nonumber \\
&& \delta_{3} = ( L - 1 )^3 \left( {9 \over 16} - {1 \over 2} \pi^2 
   - {4 \over 3} \psi^{(2)} (1) \right) , 
\label{eq:hfull}
\end{eqnarray}
where $\psi^{(n)} (z)$ is  the $n$-th order polygamma function, 
$\psi^{(n)} (z) = d^n \psi (z) / dz^n$, $\psi  (z) =
\Gamma' (z) / \Gamma (z)$. It is worth noting that computing the radiator at 
${\cal O} (\beta^3)$ requires a redefinition of the normalization in front of
the exponentiated term, $\Delta_3$, which picks up an ${\cal O} (\beta^3)$
contribution, $\delta_3$, originating from the Gribov-Lipatov form factor.  

In the following, a sample of numerical results will be shown and commented.
Only results concerning $\mu$ cross sections will be considered, since the
for\-ward-back\-ward asymmetry requires an analysis beyond the aim of the 
pre\-sent paper. 

In Fig.~1 the separate effects of the NL ${\cal O} (\alpha^2 )$ and 
${\cal O} (\beta^3)$ corrections to the radiator are shown for the QED
corrected  cross section as a function of the $s'$ cut defined as $s' / s \ge
x_{cut}$. The relative deviations with respect
to the cross section computed by means of the  radiator of 
eq.~(\ref{eq:hnt}) are shown for several centre of mass energies. Both the 
${\cal O} (\beta^3)$ (Fig.~1a) and  NL ${\cal O} (\alpha^2 )$ (Fig.~1b)
corrections amount to a contribution of several 0.1\% when the $Z^0$ radiative
return is included, but they tend to compensate one another. When the $Z^0$
radiative return is excluded, or near the $Z^0$ resonance, the 
NL ${\cal O} (\alpha^2 )$ corrections are confined at the level of 0.01-0.02\%, 
whereas the 
${\cal O} (\beta^3)$ ones remain at the level of 0.05-0.1\%. It should be
noted that in the recent analysis shown in~\cite{sm}, when studying the radiative 
corrections to
two-fermion production at LEP2, only the effect of NL ${\cal O} (\alpha^2)$ has
been taken into account, leading to results which, in the light of the present
study, are incomplete. 

Figure 2 shows the relative deviations of the cross section computed with the
full radiator of eq.~(\ref{eq:hfull}) with respect to the one computed by means
of the  one of eq.~(\ref{eq:hnt}). In Fig.~2a the effect of the newly proposed 
additive 
radiator close to the resonance is quoted. It is worth noting that at the
$Z^0$-peak the NL ${\cal O} (\alpha^2 )$ plus ${\cal O} (\beta^3)$ 
corrections introduce a systematic shift 
of about $- 0.07 $\%, dominated by the ${\cal O} (\beta^3)$ corrections.  
%which is at the same level of the present experimental error
%on the peak hadronic cross section. 
This effect has not been included in the
analyses of precision calculations performed in~\cite{cern9503}, but the
present experimental accuracy~\cite{blond} requires that it is carefully taken
into account. Going beyond the $Z^0$ peak (Fig.~2b), the effect of the radiator
of eq.~(\ref{eq:hfull}) amounts to about $ - 0.1 $\% 
when the $Z^0$ radiative return
is excluded, raising to about 0.25\% when it is included. By combining the
information of Figs.~1 and 2, one concludes that taking into account 
the NL ${\cal O} (\alpha^2 )$ corrections but neglecting the 
${\cal O} (\beta^3)$ ones, 
leads to theoretical predictions for the two-fermion processes that 
are underestimated by about 1\% in the
inclusive cases. Again, in view of the experimental precision foreseen for the
inclusive  cross sections, this effect has to be taken into account in
the theoretical predictions. Moreover, it is worth noting that in the 
case of the hadronic cross section, the effect 
of the full additive radiator of eq.~(5) as compared to the one of eq.~(3), grows up 
from 0.25\% (see Fig.~2b) to around 
0.4\% when including the $Z^0$ radiative return. 
% this case represents a realistic 
%situation,  the absolute value of the effect is comparable with the 
%foreseen experimental error, and hence the effect is phenomenologically relevant. 

We notice that, in the case of LEP1 energies, 
the results of the present analysis concerning 
NL ${\cal O} (\alpha^2 )$ and ${\cal O} (\beta^3)$ corrections confirm the 
estimates obtained in~\cite{apl}, with the conclusion that, whereas at that time 
such effects were marginal, at present, and in the light of the continuous 
progress in the reduction of the experimental errors, they are significant.  
As far as the LEP2 regime is concerned, the present analysis points out the relevance 
of IS higher order hard photon effects, and completes a first attempt to 
quantify such contributions recently performed in~\cite{sm}.  

As already pointed out above, in~\cite{berends} a catalog of the radiators 
available at the time of the LEP1 workshop is given, and  several 
options are offered, according to the particular form of the radiators and 
the contributions taken into account.  In particular,  the options can be 
grouped as follows: options (A) and (B) refer to the radiators reported in 
the present  paper in eqs.~(4) and (3) respectively, and are in the additive 
form;  options (C), (D) and (E), on the contrary, describe radiators in the 
factorized form, and are superseded by the radiator described in~\cite{jsw,apl}. 
The radiator presented here, eq.~(5), is intended to supersede options (A) and 
(B), i.e. the additive radiators at present used in some computational tools 
for the LEP analysis. Given this situation, it is worth at this point to 
examine carefully in which relation the best additive and factorized radiator 
are. To this aim, in Fig.~3 a detailed comparison of the radiators in themselves
as functions of their argument (Fig.~3a) and of the corresponding cross sections 
(Fig.~3b) as functions of the invariant mass cut after ISR  
is shown. The comparison of the radiators shows that their relative difference 
is within $3 \times 10^{-4}$ for most of the spectrum, raising up to  
$3 \times 10^{-3}$ in the very hard tail. Analogously, the comparison of the 
cross sections shows that their relative difference is contained within 
$2.5 \times 10^{-4}$ for centre of mass energies ranging from LEP1 to LEP2, and 
for  $ 0.01 \le x_{cut} \le 0.99 $, i.e. for every realistic situation. From 
the present analysis, one can conclude that the two radiators are equivalent
from the phenomenological point of view. 

The numerical results shown  in  the present  paper point  out that the additive  
radiators implemented in standard computational  tools used for  the  two-fermion  
data analysis at LEP have to be upgraded both for  LEP1 physics, where they do not 
fulfil the present  precision requirements, and for LEP2 physics, where they do not  
describe appropriately the $Z^0$ radiative return. This  is a first result  of the 
present analysis. Moreover, given  the results  shown in Fig.~3,  one can conclude  that 
the upgrade can equivalently be performed either by definitely substituting the  additive  
radiators by means of the factorized  one  of ref.~\cite{jsw,apl}, or  by simply modifying 
the ones already implemented with the  addition of  a few  FORTRAN lines, according to the
newly proposed radiator of eq.~(\ref{eq:hfull}). This is a second  result of the present
analysis.  At this point, 
the choice is  a matter of taste/convenience.\footnote{\footnotesize In the approach 
of  ref.~\cite{prd}, the ``soft  photon singularities'' are regularized by means  of the 
variance-reduction technique known as ``control variates''; for its actual implementation, 
besides the radiator/structure function  one needs  also their  primitives,  possibly in
analytical form in order to save CPU time; hence, for practical 
reasons, the   implementation of an additive radiator  is  preferable. }     

Summarizing, the  ${\cal O} (\beta^3)$  
contribution to the QED additive radiator has been 
considered, and a new additive radiator, accounting for 
NL ${\cal O} (\alpha^2 )$ and ${\cal O} (\beta^3)$ corrections, is proposed. 
The effect of the ${\cal O} (\beta^3)$ correction on two-fermion cross sections 
at  LEP1 and  LEP2 has been  investigated  and critically
compared with the effect of   NL  ${\cal O} (\alpha^2)$ 
corrections. The  ${\cal O} (\beta^3)$  corrections 
turn out to be relevant in view of the experimental precision already 
reached at LEP1; moreover, together with the  NL  ${\cal O} (\alpha^2)$ terms, 
they represent a non-negligible contribution to ISR because of the $Z^0$ radiative return  
and in view of the foreseen experimental precision at LEP2. 
The newly proposed  additive radiator improves considerably 
the ones implemented in standard computational tools for two-fermion processes 
at LEP. 

%\newpage
\bibliography{bank}

\begin{figure}[hbtp]
\begin{center}
\epsfig{file=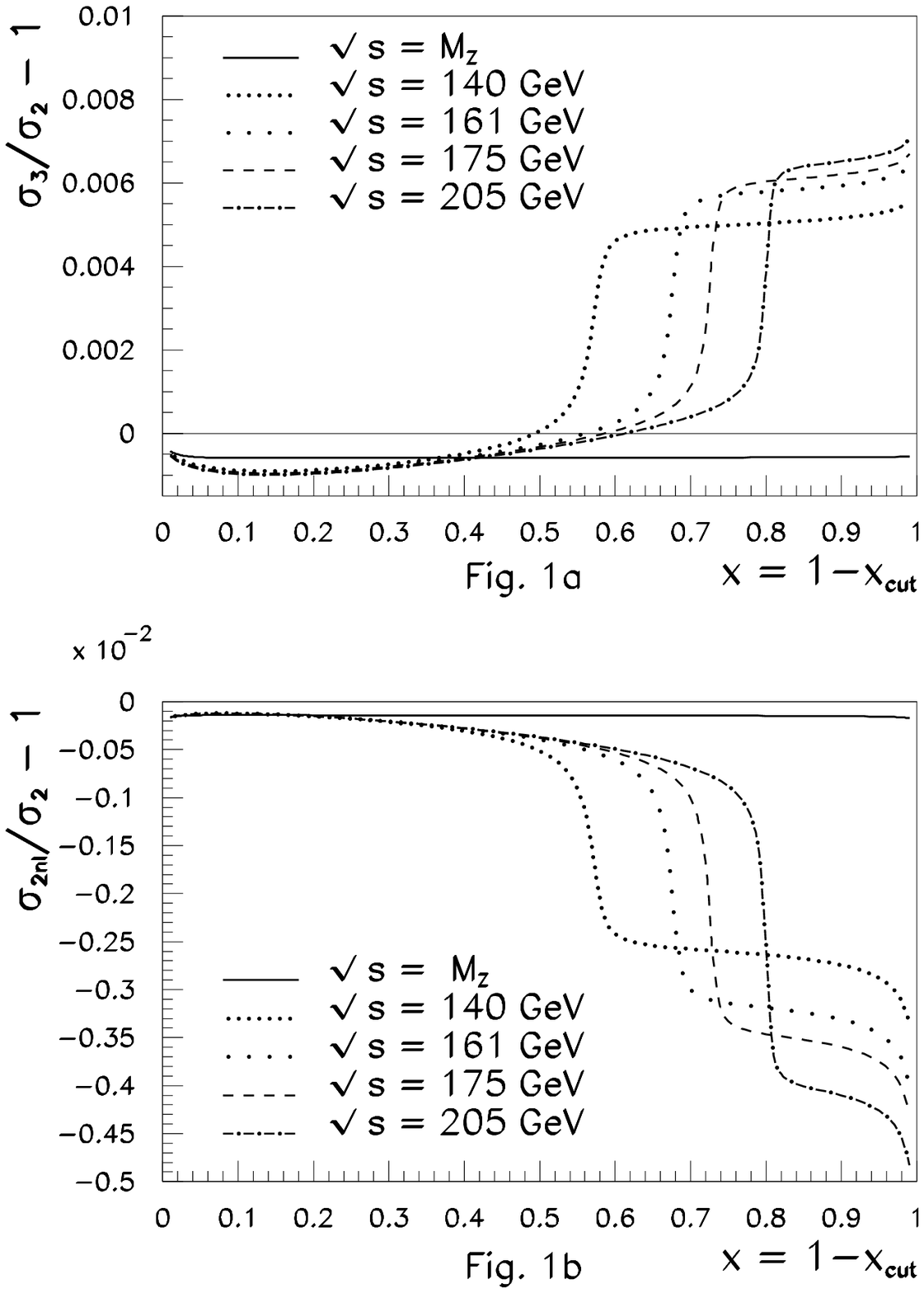,height=16truecm}
\end{center}
\caption{The relative deviation between the cross sections computed by means of
the radiators of eq.~(5) with the substitution $\delta_2^H (1-x,s) \to h_2
(x,s)$ ($\sigma_3 $ in a)  
and  of eq.~(4) ($\sigma_{2nl}$ in b),  and the radiator of eq.~(3)
($\sigma_2$), as a function of the invariant mass cut.  } 
\label{fig:fig1}
\end{figure}

\begin{figure}[hbtp]
\begin{center}
\epsfig{file=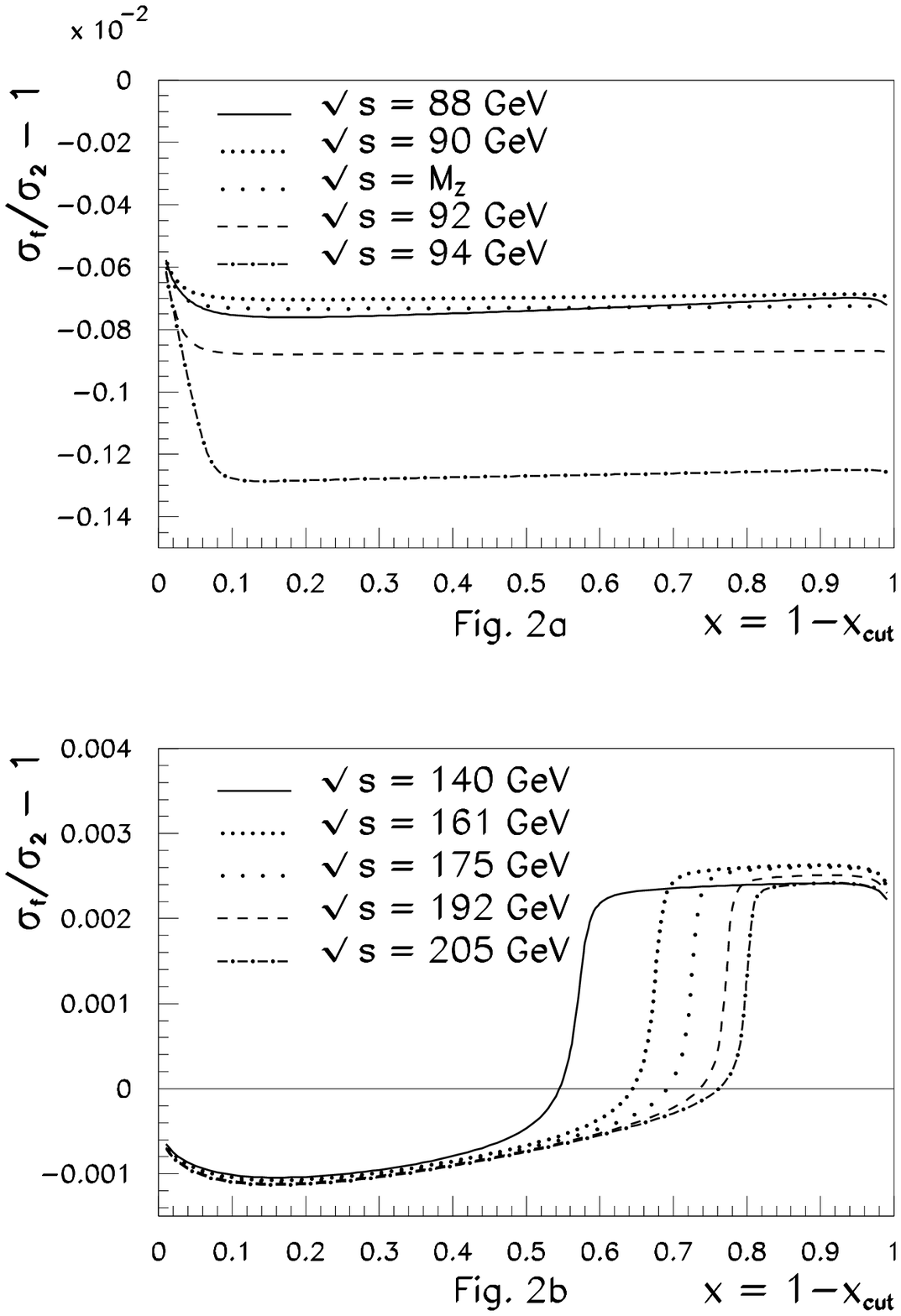,height=16truecm}
\end{center}
\caption{The relative deviation between the cross sections computed by means of
the radiators of eq.~(5) ($\sigma_f $) 
 and the radiator of eq.~(3)
($\sigma_2$), as a function of the invariant mass cut. The LEP1 (a) and LEP2
(b) cases.  } 
\label{fig:fig2}
\end{figure}

\begin{figure}[hbtp]
\begin{center}
\epsfig{file=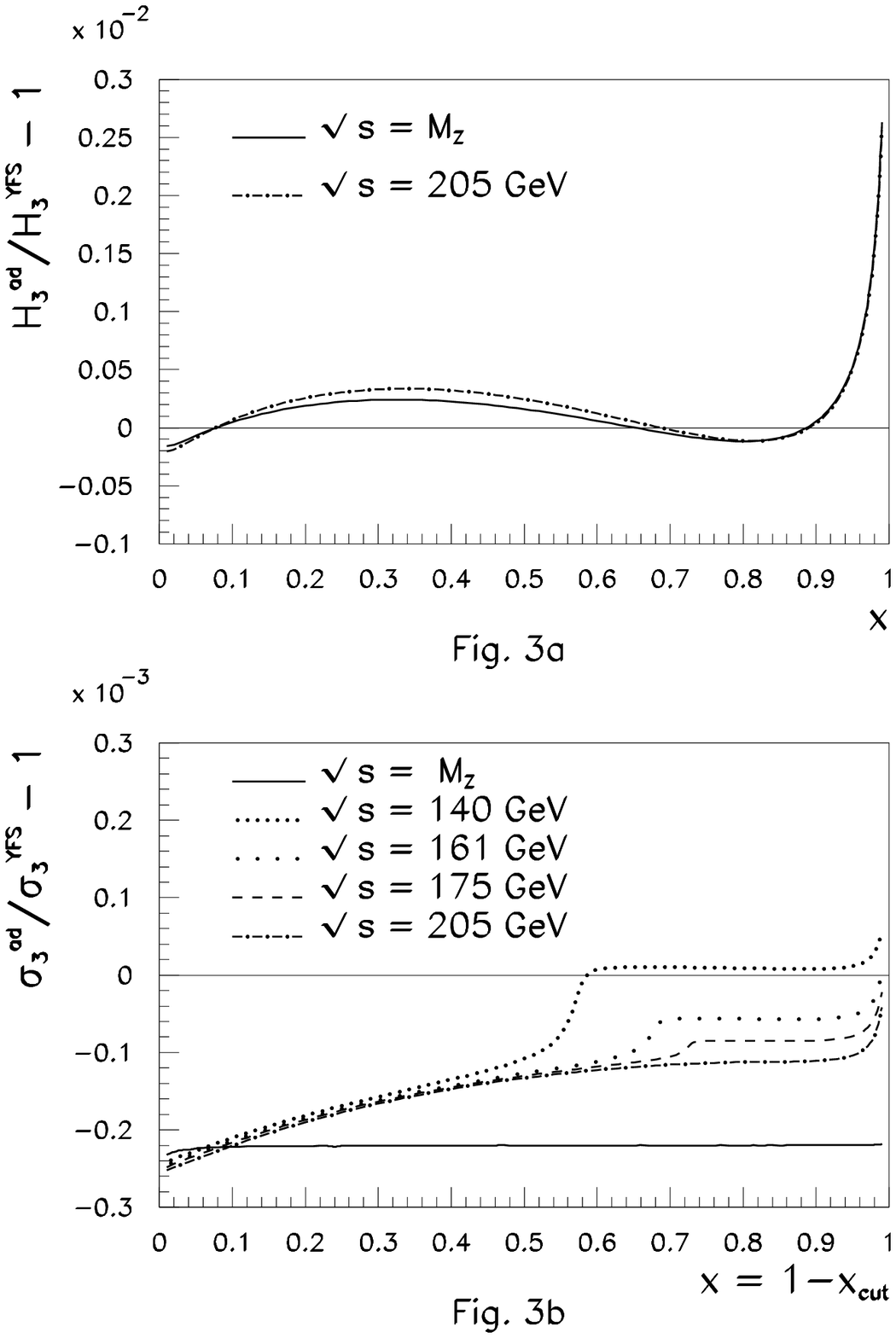,height=16truecm}
\end{center}
\caption{The relative deviation between the radiators of eq.~(5) $H_3^{ad}$  
and of [4] $H_3^{YFS}$ (Fig.~3a), and the corresponding cross sections 
(Fig.~3b).} 
\label{fig:fig3}
\end{figure}

\end{document}